# Testing Outlier Detection Algorithms for Identifying Early-Stage Solute Clusters in Atom Probe Tomography

R S. Stroud[1], A. Al-Saffar[1], M. Carter[2], M P. Moody[2], S. Pedrazzini[1], M R. Wenman[1]

1 Department of Materials and Centre for Nuclear Engineering, Imperial College London, SW7 2AZ, UK

2 Department of Materials, Parks Road, Oxford, OX1 3PH, UK

## Abstract

Atom probe tomography is commonly used to study solute clustering and precipitation in materials. However, standard techniques, such as the density based spatial clustering applications with noise (DBSCAN) perform poorly with respect to small clusters of less than 25 atoms. This is a fundamental limitation of density-based clustering techniques due to the usage of $N_{min}$, an arbitrary lower limit placed on cluster sizes. Therefore, this paper attempts to consider atom probe clustering as an outlier detection problem of which KNN, LOF, LUNAR algorithms were tested against a simulated dataset and compared to the standard method, for a range of cluster sizes. The decision score output of the algorithms was then auto thresholded by the Karcher mean to remove human bias. Each of the major models tested outperforms DBSCAN for cluster sizes of less than 25 atoms but underperforms for sizes greater than 30 atoms. However, the new combined $k$NN and DBSCAN method presented was able to perform well at all cluster sizes. The combined $k$NN and DBSCAN method is presented as a possible new standard approach to identifying solute clusters in atom probe tomography.

## 1. Introduction

In many materials, clusters of a few atoms can occur. Such as in irradiated steel, Zr alloys and Al[1]–[3]. However, detecting very small clusters of fewer than 30 atoms is problematic using current techniques such as density-based spatial clustering of applications with noise (DBSCAN) and the maximum separation method (MSM). They are also prone to errors from human bias [4]. For irradiated alloys, neutron irradiation creates point defects comprising of vacancies and interstitial atoms, as incoming neutrons knock atoms out of their lattice sites. Vacancies can then act as nucleation sites for the formation of solute clusters.

In reactor grade ferritic steel, the solutes, most notably Mn, Ni, Si, Cu and P, over time at temperature, cluster around point defects via diffusion into a more energetically favourable configuration. There is a strong association between solute clusters formed by neutron irradiation damage in reactor steel and their mechanical properties [5] and one source is likely due to the solute clusters acting to prevent or hinder the movement of dislocations. In turn, the restricted dislocation motion contributes to an increase in the material's strength and brittleness. The cluster volume fraction has been used to develop embrittlement trend curves [6]. The embrittlement trend curve forms a foundation for evaluating the structural integrity of a reactor pressure vessel (RPV) and is used for reactor life extension. Currently, atom probe tomography (APT) is used to identify and analyse solute clusters in RPV steel due to the 3D nature of clusters and the innate ability of APT to characterise the size and composition of clusters.

DBSCAN is a density-based method that works by the user setting two key parameters $D_{max}$ and $N_{min}$. An atom is defined as a core atom if there are at least a minimum number of atoms, $N_{min}$, within a distance $D_{max}$. In contrast, an atom is considered a border atom if it is within a distance of $D_{max}$ to a core atom but less than $N_{min}$ within a distance $\epsilon$ to itself. All other atoms are not part of the atomic cluster. DBSCAN has two key advantages that make it a good choice for identifying clusters of atoms in APT; it is more sensitive to smaller clusters than voxel-based approaches (for example using isoconcentration surfaces) and it can determine the number of clusters. Whilst DBSCAN, is more sensitive to smaller clusters than voxel-based approaches, it is unable to distinguish between small clusters and noise introduced from the limitations of APT. Consequently, early-stage, small-scale (10s of atoms) clusters, are inevitably understudied because they are difficult to isolate. However, small-scale embryonic clusters could provide valuable insights into the later formation of larger clusters. Improvements in APT resolution, detection efficiency and cluster identification methods would allow for a more comprehensive study into atomic scale cluster formation.

Alternative methods for cluster identification have been proposed, such as ordering points to identify the clustering structure (OPTICS) [7] and the isoposition method (IPM) [8], [9]. However, there is no widely accepted replacement for DBSCAN that provides sensitivity improvements to smaller clusters and provides a reduction in human input parameter sensitivity. This is an area where the atom probe community has shown from round robin studies [4], [10] that there can be huge discrepancies for identical atom probe data sets.

Machine learning and neural networks have been proposed for APT analysis [11], [12]. However, deep learning methods have traditionally fallen behind statistical approaches, for unsupervised learning and clustering. Neural networks are machine learning models that approximate a function by using a network of nodes whereby values propagate through, being multiplied by a series of weights between each layer. The loss (error) of the network is calculated against a training set of values and minimized through an optimization function, such as gradient descent. Deep neural networks are characterised by more than one hidden layer between their input and output layers. The additional hidden layers enable a composition of features resulting in the ability to model more complex functions [13], [14]. However, the additional layers increase the computational requirement and require extra architectural features to train accurately [15], [16]. The gradient of the loss function is used to determine the change in parameters needed to learn.

Recently it has been shown that deep learning models can outperform the previous standards of DBSCAN, $k$-nearest neighbours ($k$NN) [17] and the local outlier factor (LOF) [18] for a selection of clustering datasets [19]. This paper aims to apply many outlier detection models from the pyOD python outlier detection library [20]. This includes the learnable unified neighbourhood-based anomaly ranking (LUNAR) [19] and a variational autoencoder (VAE) [21], [22] for APT cluster identification to improve upon MSM and DBSCAN in increasing sensitivity and reducing human bias and error.

We propose that an atom probe dataset containing solute clusters can be considered as a 3-dimensional tabular dataset that consists of outliers and inliers. The LUNAR model as described by Groodge et al. [19] is a graph neural network (GNN) for detecting outliers. The 3-dimensional point cloud data is used to construct a $k$-nearest neighbour ($k$-NN) graph where each atom becomes a graph node with a feature vector equal to the distances to the $k$-NNs, as given in eq. (1) . As the graph is self-constructing, no user input is required to determine optimal parameters, such as $D_{Max}$ and $N_{Min}$. The output data is in the form of an outlier score that can be used to identify clusters. The VAE model uses the reconstruction error from an autoencoder to determine outlier points. An autoencoder is a neural network that takes in an input of $n$ dimensions, reduces it to a smaller than $n$ dimension latent space with an encoder, then uses a decoder to return an output with $n$ dimensions. The VAE attempts to fully replicate the original data with fewer dimensions. This results in a loss of information, and an error when reconstructing the original data. The error per point can be considered as the outlier score.

$$x_{features} = [d_1, d_2, \ldots, d_n] \quad (1)$$

## 2. Methods

### 2.1. Experimental and Simulated Data

Two neutron irradiated specimens of SA508 class 3 RPV steel (composition shown in Table 1) were provided by EDF. The pair of surveillance samples from a PWR consists of a baseline unstrained sample and a sample uniaxially pre-strained to 5% nominal strain.

Table 1: The composition of alloy SA508 class 3 forging in wt%. The data was taken from the manufacturing records.

| Element | SA508 class 3 Forgings wt. % Max |
|---|---|
| Carbon | 0.200 |
| Manganese | 1.590 |
| Molybdenum | 0.680 |
| Nickel | 0.850 |
| Sulphur | 0.008 |
| Phosphorus | 0.008 |
| Silicon | 0.300 |
| Chromium | 0.150 |
| Copper | 0.080 |
| Vanadium | 0.010 |
| Antimony | 0.008 |
| Arsenic | 0.015 |
| Cobalt | 0.020 |
| Tin | 0.010 |
| Aluminium | 0.045 |
| Hydrogen | 1ppm |
| Iron | Balance |

For irradiation, the samples were placed into a surveillance capsule in a PWR for approximately 18 effective full power years. The capsule received a fluence of 49.75 x $10^{18}$ ($E > 1$ MeV) n/cm$^2$ equivalent to a displacement per atom (dpa) of approximately 0.1.

The samples were prepared for the APT, at the Materials Research Facility (MRF) using a non-rotary semi-automatic grinder and polishing with a non-aqueous diamond polishing solution. The samples were ground with 2000 and then 4000 grit paper with 22 N of force for 3 minutes. For polishing, the samples were polished with 3, then 1, then ¼ µm alcohol-based diamond solution with 22 N of force for 3 minutes at each step. The APT needles were prepared using the Helios focused ion beam (FIB) at MRF. Two regular cross sections of (24 x 9 x 3) µm at 9.3 nA beam current, plus one regular cross section of (10 x 9 x 3) µm at 9.3 nA beam current, plus a platinum deposition of (20 x 1.5 x 0.5) µm at 0.23 nA were used for the liftout. For the APT tip sharpening four annular milling steps were used, as shown in Table 2.

*Table 2: The FIB milling procedure for APT tip preparation.*

| Step Number | Beam Current nA | Outer Radius µm | Inner Radius µm | Beam Voltage kV |
| --- | --- | --- | --- | --- |
| 1 | 2.5 | 7 | 2.2 | 30 |
| 2 | 0.79 | 2.4 | 0.7 | 30 |
| 3 | 0.04 | 1.5 | 0.1 | 30 |
| 4 | 0.19 | 1.5 | 0.1 | 2 |

The APT was carried out using the UK National Nuclear Users Facility active atom probe at the University of Oxford, which is a Cameca LEAP 5000XR. The two samples presented were run in laser mode (S2) and voltage mode (S1). The laser mode was run at a temperature of 50 K, a laser energy of 50 pJ, a detection rate of 0.2% and a pulse frequency of 200 kHz. These parameters were chosen to keep in line with similar previous experiments [23]. The voltage mode was run at a temperature of 55 K, a detection rate of 0.2% and a 20% pulse fraction. The parameters were chosen to maximize yield in lieu of a co-fracturing event and to maintain similar parameters with previous experiments [24]. An additional sample with more pronounced clustering due to a lower temperature (~155°C) and higher fast neutron flux ($9 \times 10^{12} ncm^{-2}s^{-1} \pm 10\%$) of irradiation, S3, was also studied. This sample data was previously published in Carter et al. [24].

The raw data was reconstructed using APSuite 6.1. The initial tip radius for the reconstruction was selected by using the detector heat map to identify low density regions that correspond to crystallographic poles. Subsequently, the reconstruction of the position coordinates (*x*, *y*, *z*) and the mass to charge ratio (*mc*) were exported from APSuite in a comma separated variables (csv) file format. The csv position data and the range file were imported into python. Each element was filtered based on the range data. The final data used for analysis consisted of a Fe dataset and a major solutes dataset. The models were only applied to the solutes, in a similar fashion to the MSM and DBSCAN. For this type of low alloy bainitic steel this is a straightforward decision as Fe makes up over 95% of the material. However, for other materials, this decision is another input parameter, i.e. which elements are to be selected.

### 2.2. Model Selection

A variety of models have been proposed and tested for outlier detection since DBSCAN. However, most are not appropriate for APT cluster analysis as they require knowledge of the number of clusters beforehand, whilst this information is not available in APT datasets. In addition, most methods do not improve upon the sensitivity and the precision of DBSCAN to a significant level. The models, which are being tested in this paper, may overcome these two challenges. Like DBSCAN, the number of clusters are not needed before analysis. This paper will focus primarily on LUNAR and *k*NN as these models had the best performance. In addition, a new method is proposed, a combined *k*NN and DBSCAN hybrid method.

## 2.3. Model Architecture
### 2.3.1. LUNAR
The network architecture was optimised to improve performance on the simulated data. The resulting neural network used consists of 15 hidden layers of 32 nodes using the Gaussian error linear unit (GeLU) activation function for all layers, as shown in figure 1. Every second layer was a batch normalisation layer, normalising the output of the dense layers. Batch normalisation accelerates learning by reducing covariate shift and mitigating the vanishing/exploding gradient problem [25], [26]. The forward pass is shown in equations (2)-(5), where $x$ is the initial input or the nearest neighbour distances $d_n$. The initial *out* is the sum of $\alpha \cdot x$, which is a weighted average of the output weights $\alpha$ and the input nearest neighbour distances $x$. The final *out* given in eq. (5) is normalised with a sigmoid function, whilst the output ranges between 0 and 1, this output does not representant a probability.

The GeLU activation function was selected due to the increase in performance, which was shown by Hendrycks and Gimpel [27]. The model was built using the pyOD outlier detection library [20]. In addition, LUNAR was implemented as an average ensemble of three LUNAR models with different random seeds.

$$\alpha = \text{net}(x) \qquad (2)$$
$$\alpha = \text{softmax}(\alpha) \qquad (3)$$
$$out = \text{sum}(\alpha \cdot x) \qquad (4)$$
$$out = \text{sigmoid}(out) \qquad (5)$$

### 2.3.2. VAE
The network architecture consists of an encoder containing three layers with hidden sizes of 64, 32, 8, and a decoder containing three layers with hidden sizes of 8, 32, 64. The VAE had a dropout of 0.2, a latent dimension of 2, and a beta value of 50. Dropout is a regularization technique that reduces the chance of overfitting, it sets a random fraction of input units to zero at each training step [28]. The latent dimension is the number of neurons at the narrowest part of the autoencoder and the beta value is the regularization term applied to the loss function. It biases the latent distribution to a gaussian distribution, enabling a more disentangled latent space. This disentanglement reduces the reconstruction accuracy but improves performance when using the VAE for outlier detection [21].

## 2.4. Network Training
### 2.4.1. LUNAR
Each APT sample varies substantially, in terms of its features, artefacts and structure. To maximize performance, the model is trained on each sample independently. The network itself is small at 19,073 parameters to keep the training step fast enough to be run per APT sample. For training a 16 core Xeon W-2245 @ 3.9GHz was used. The training and evaluation took approximately 1 minute per 3 million ion APT sample, where each model is trained on a subset of each sample before evaluating on the entire sample.

The optimization algorithm used was AdamW [29], with a weight decay of 0.1 to reduce overfitting, and a $\beta_2 = 0.95$ for improved generalization, a learning rate of $1 \cdot 10^{-4}$ and 50 epochs were used.

To generate a mixed labelled dataset for training, a third of the atoms are sampled at random and new atoms are generated a small distance $\epsilon$ away (as shown in Figures 1a to 1b) and another sample of randomly generated atoms introduced. These two methods for generating outlier labels are mixed in equal proportion. These atoms are then considered outliers as they are not part of the original dataset. A *k*-nearest neighbors (*k*-NN) graph is created. Each atom is a node connected to its *k*-nearest neighbours; the distances are used as the inputs for the graph neural network (GNN), the input is shown in equation (1) and in Figure 1c. The model outputs an array of weights, a weight for each input distance, which are then used for a weighted average of the nearest neighbour distances as shown in eq. (6). A simpler approach would set all the weight values to 1, as shown in eq. (7). This results in a simple average of all the nearest neighbour distances; this is the definition of *k*NN. Here, we have deviated from the original implementation of LUNAR to be able to adaptively learn on unseen features, which are commonplace in APT datasets. LUNAR was used as a part of an ensemble of three LUNAR models, each trained with a different random seed to improve the overall F1 score performance on the test dataset. Ensemble learning is a commonly used technique to increase the generalizability of learning [30].

When training LUNAR on known APT datasets, it is unfeasible to identify unseen features in new data. For example, when LUNAR is trained on a dataset containing solute clusters, it is not possible to identify a solute segregated grain boundary in a different sample. This could be overcome by using a large and varied training dataset. However, this is not available at this time, given the difficulty of obtaining the irradiated steel data and the extreme variety between samples. Therefore, LUNAR, in this paper, was trained on a subset of samples to be analyzed. This results in the ability to identify all features found in irradiated RPV steel, which current MSM and DBSCAN cannot do. For comparison, we have included the simple average approach in our analysis. This *k*-nearest neighbours (*k*NN) approach has been well documented [31]. The hypothesis is that *k*NN and LUNAR will perform similarly, but LUNAR may perform better with more complex samples and for a larger range of *k* values due to the adjusting weights.

$$[d_1, d_2, d_3, \ldots, d_n] \rightarrow \text{LUNAR} \rightarrow [w_1, w_2, w_3, \ldots, w_n] \quad (6)$$
$$\rightarrow \text{Sigmoid}(\text{Sum}(d_1 w_1, d_2 w_2, d_3 w_3, \ldots, d_n w_n)) \rightarrow \text{LUNAR\_Score}$$

$$[d_1, d_2, d_3, \ldots, d_n] \rightarrow \text{LUNAR} \rightarrow [1,1,1,\ldots] \rightarrow \text{Average}(d_1, d_2, d_3, \ldots, d_n) \rightarrow \text{KNN\_Score} \quad (7)$$

LUNAR trained per APT sample with deep network results in a high likelihood of overfitting. To overcome this, a dropout of 20% per layer was used.

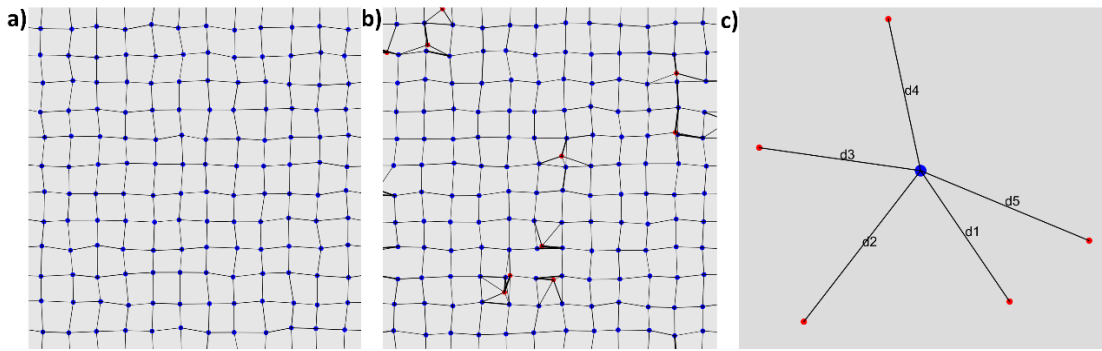

*Figure 1. 1a is an illustration of a 2D lattice showing a representation of the original data. 1b is an illustration of the added atoms in red that are labelled as outliers. The lines connect the nearest 5 neighbours, and the line thickness corresponds to the distance. 1c is an illustration of a single node and its associated connections labelled d1 to d5. A vector of distances is used as the model input.*

### 2.4.2. Training of *k*NN, VAE and LOF Models

The three remaining models, *k*NN, VAE and LOF were trained similarly to LUNAR. Each model was trained on each APT sample before outputting predictions on the same sample. All three models were trained with a range of numbers of nearest neighbours' parameters of $k = 2$ to $n = 100$. For the simulated data, all models performed best at $n = 10$. The *k*NN model used the nearest neighbour mean average distance. The VAE used the original atomic position coordinates plus the $n$-nearest neighbour distances as inputs. For both *k*NN and LOF, the negative of the decision scores were used.

### 2.4.3. A *k*NN and DBSCAN Hybrid

To attempt to capture the best performing model across all regimes, a hybrid model was also tested. The *k*NN and DBSCAN hybrid model is defined as the average distance between the *k* nearest neighbours (as in *k*NN) multiplied by the binary output from DBSCAN. The standard output from DBSCAN consists of a 1 for an atom considered to be in a cluster and 0 for an atom not in a cluster. To compose *k*NN and DBSCAN, DBSCAN's outputs were changed to 0.5 for a cluster atom and 1.5 for a non-clustering atom. This new multiplied output score will follow the *k*NN scores for small cluster sizes, as DBSCAN will return the same output for each atom. For example, if $N_{min}$ is set to 20, if there is a cluster of size 15, DBSCAN will return a multiplier of 1.5 for all atoms. Therefore, not affecting the output in any meaningful way, any variation in scores will have come from *k*NN. However, if the cluster has a size of 25, then most of the variance in output scores will be a result of DBSCAN. For the DBSCAN component, the parameters were set at a $D_{max}$ and $N_{min}$ of 0.8 and 15 respectively. This was chosen to be a representative selection for the optimal parameters for the higher cluster sizes.

### 2.5. Clustering Threshold

The output from all four models is a single number representing an outlier score. The output outlier score is different for each model. For *k*NN, the output is simply the average distance from an atom to its nearest neighbours. For LUNAR, the outlier score is the weighted average distance from an atom to its nearest neighbours, where the weights are decided by the model. For VAE, the outlier score is the reconstruction per atom as the squared distance from the original atom. For LOF, the outlier score is the local density. This is determined by the using reachability distance divided by the number of neighbours. Where the reachability distance between point A and point B, is the distance between point A and the $k^{th}$ nearest neighbour of B. This method produces a value that is greater than 1 when the density is greater than the local density and less than one for the inverse. This density value can then be used in a similar way as DBSCAN and *k*NN, whereby a higher density of solutes indicates the presence of a solute cluster.

This continuous value output allows for additional analysis, which is not possible using DBSCAN. However, if a binary output is needed, some threshold on the outlier score is required (where a score above the threshold is considered as an atom belonging to a cluster and a score below the threshold is considered as an atom not in a cluster). To determine the threshold value, a statistical method is needed. Firstly, a Gaussian distribution with a threshold value set at some standard deviation away from the mean was used. However, initial tests showed this method performs poorly. This is largely due to the distribution of scores not fitting a Gaussian distribution. In addition, the score histograms greatly vary with each sample, resulting in needing to use multiple standard deviation values. Instead, a Karcher mean or Riemannian manifold centre of mass was preferred. The Karcher mean is the same as a Euclidean mean in flat space but differs in curved space. To find an appropriate threshold, the Karcher mean plus one standard deviation (Riemannian version of standard deviation) was used. The use of an automatic thresholding method implies that all four models being tested can be used without any user (human) input, unlike DBSCAN or MSM. In addition, an approach similar to DBSCAN for APT can also be used, using the score distribution histogram for the sample to select a cut-off that ensures no random atoms are labelled as clusters. However, this adds an additional step (two additional steps for DBSCAN), whereby operator bias is introduced.

## 2.6. Model Evaluation

To evaluate the performance of these models, a simulated dataset was created. The simulation is created using a python factory class that takes as initial parameters: the number of unit cells per side, the cluster relative density and the cluster atom counts. From these parameters, a simple BCC steel lattice is created with a composition equal to 0.81 Ni, 1.62 Mn, 0.07 Cu and 97.50 Fe (at%). This was chosen to keep the solute composition similar to the manufacturing records of the steel. The clusters are then created and added to the lattice by selecting random coordinates within the cell boundary. A normal distribution around this point and the selection for the initial cluster density determine the positions of the solute atoms. Solute atoms that were previously inside this newly defined cluster are removed to provide a clean dataset for the model. Atoms without an element assigned are placed around the random point. They are then assigned an element, based on the specified composition of the material. To simulate the running of a sample in APT, 45% of the atoms were removed at random and each remaining atom was perturbed by a noise value of $x = 0.5\ nm, y = 0.5\ nm, z = 0.1\ nm$ as seen in Figure 2. These noise values were taken from those used by Hyde et al. [8]. Each model was also tested using a range of relative cluster densities. The relative cluster density is defined as cluster density relative to the solute matrix. As the matrix of solutes is sparse, high relative densities were required.

All models: LUNAR, *k*NN, LOF, VAE, DBSCAN and the *k*NN/DBSCAN hybrid were then compared on the simulated data using the F1 score. The F1 score is defined as the mean of the precision and the recall, where the precision is defined as the true positives to total positives ratio, and the recall is defined as the ratio of true positives to the sum of true positives and false negatives. The models were tested on three simulated datasets made with different random seeds.

The F1 score was used instead of the area precision recall curve, due to the difficulty in producing a fair comparison with DBSCAN. DBSCAN produces a binary output making the F1 score a more accurate comparison. For LUNAR, KNN, LOF, and VAE, the F1 score was calculated from the threshold selected by the Karcher mean. In the case of DBSCAN, a parameter search was performed by sweeping through the combinations of $D_{max}$ and $N_{min}$. First, the maximum $N_{min}$ is defined at 30 and the $D_{max}$ resolution is defined at 0.05. For each value of $N_{min}$, the minimum of value of $D_{max}$ that produces more than zero clusters is selected. After this parameter sweep, an F1 score for every value of $N_{min}$ is calculated. The top 5 highest F1 scores are averaged into a single F1 score for DBSCAN. This approach provides similar results to a full parameter sweep, whilst maintaining fairness and a reasonable computation time.

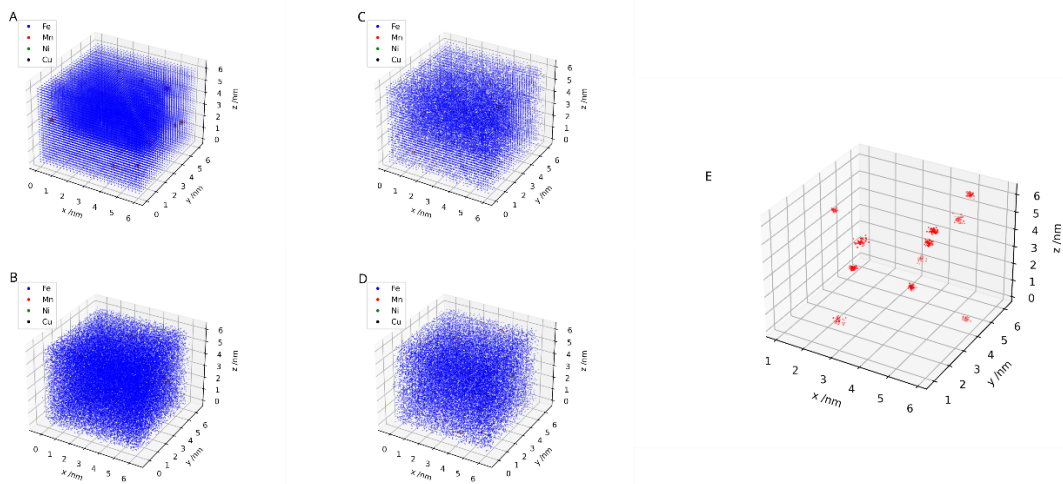

Figure 2: This represents the simulation output of the BCC Fe lattice with the two steps to introduce noise and atom removal. Each figure is a newly generated dataset. A is data with no noise or the removal of atoms. B is a dataset with noise added. C is a dataset with removal of 45% of atoms. D is a dataset with both the added noise and the removal of atoms. E is a dataset with added noise and the removal of atoms, with the Fe removed (solute clusters shown in red).

## 3. Results
### 3.1. Simulated Data

Figure 3 shows that both LUNAR and *k*NN outperformed DBSCAN on simulated data for cluster sizes smaller than 25 atoms. It was decided to remove all F1 scores below 20% from the results. This represents the random guess marker, below which the concept of cluster detection in atom probe starts to break down.

For clusters greater than 30 atoms in size, LUNAR and *k*NN performs worse than DBSCAN. However, in this test case, the new *k*NN/DBSCAN hybrid model seems to capture the small cluster size performance of *k*NN, whilst also reverting back to DBSCAN performance at larger cluster sizes. In addition, all other unsupervised models tested performed worse than LUNAR, *k*NN and the hybrid model.

The relative densities between 100 and 800 showed no noticeable difference in F1 score for any of the models as shown in Figure 4. At a relative density of 50, all models perform worse than relative densities above 100. The difference in F1 score heatmaps in Figure 5 (where positive numbers/blue colours indicate improved performance) also confirms that *k*NN and LUNAR perform better for smaller clusters around 20 or fewer, irrespective of the cluster densities.

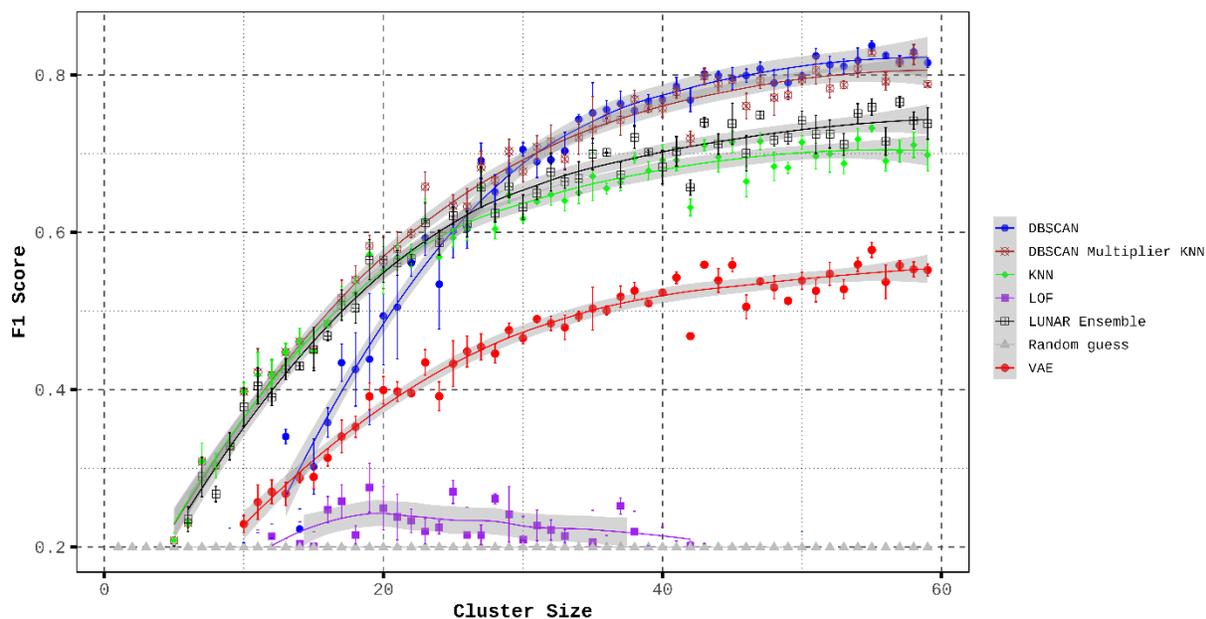

*Figure 3: A Comparison of six different clustering models, using F1 score as the classification score. LOF (purple) and VAE (red) performed worse than DBSCAN (blue). LUNAR (black), DBSCAN/ kNN hybrid (brown) and kNN (green) outperform DBSCAN for small cluster sizes. A random guess is highlighted at a F1 score of 20% in grey. The y-axis is cut at the random guess marker of 20%. The error bars are the standard errors from three runs.*

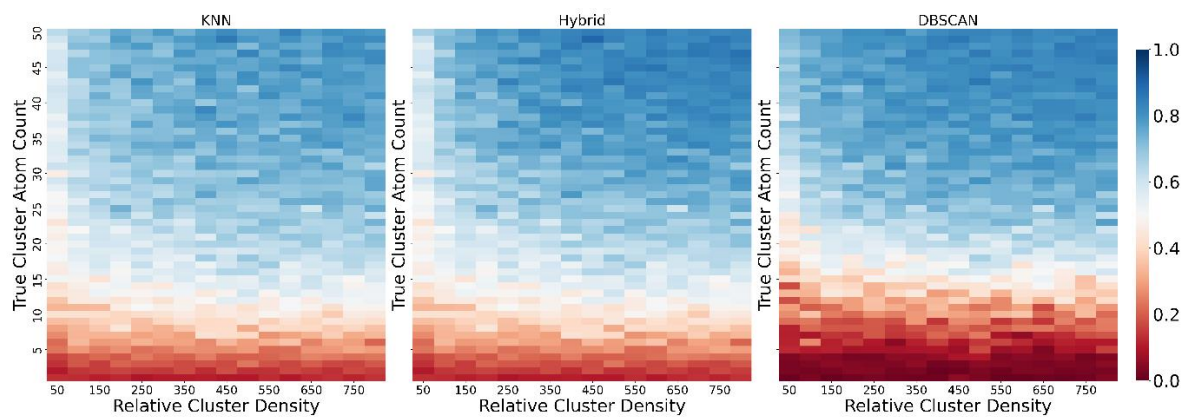

*Figure 4: F1 score Heatmaps for DBSCAN, kNN, and LUNAR for each cluster atom count and relative cluster density pair.*

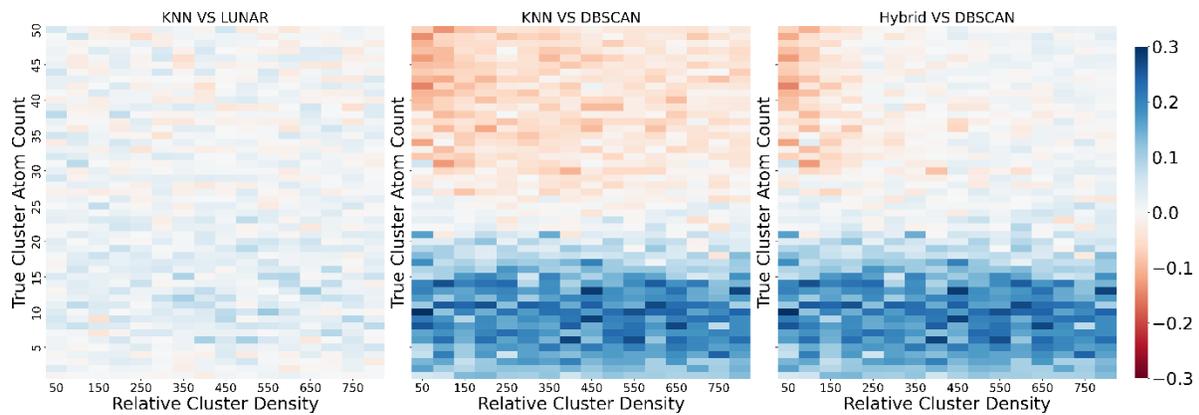

*Figure 5: Difference in F1 score heatmap comparisons of each pair of DBSCAN, kNN, and LUNAR for each cluster atom count and relative cluster density pair. The negative values (red colour) correspond to the second model scoring higher than the first, the positive values (blue colour) indicate the inverse.*

Moreover, *k*NN and LUNAR were tested at different values of $k$. The results (Figure 6) shows that LUNAR is less sensitive to $k$ than *k*NN. At *k = 10*, the blue coloured heatmap shows *k*NN scoring approximately 30% higher overall. The inverse is true for *k = 100*, where LUNAR is scoring approximately 10% higher overall. This is likely due to the ability of LUNAR to adapt the output weights to compensate for an incorrect $k$. However, if $k$ is selected correctly, then *k*NN outperforms LUNAR. For the simulated data the ideal *k* value is 10. However, this ideal value will change for each dataset analysed.

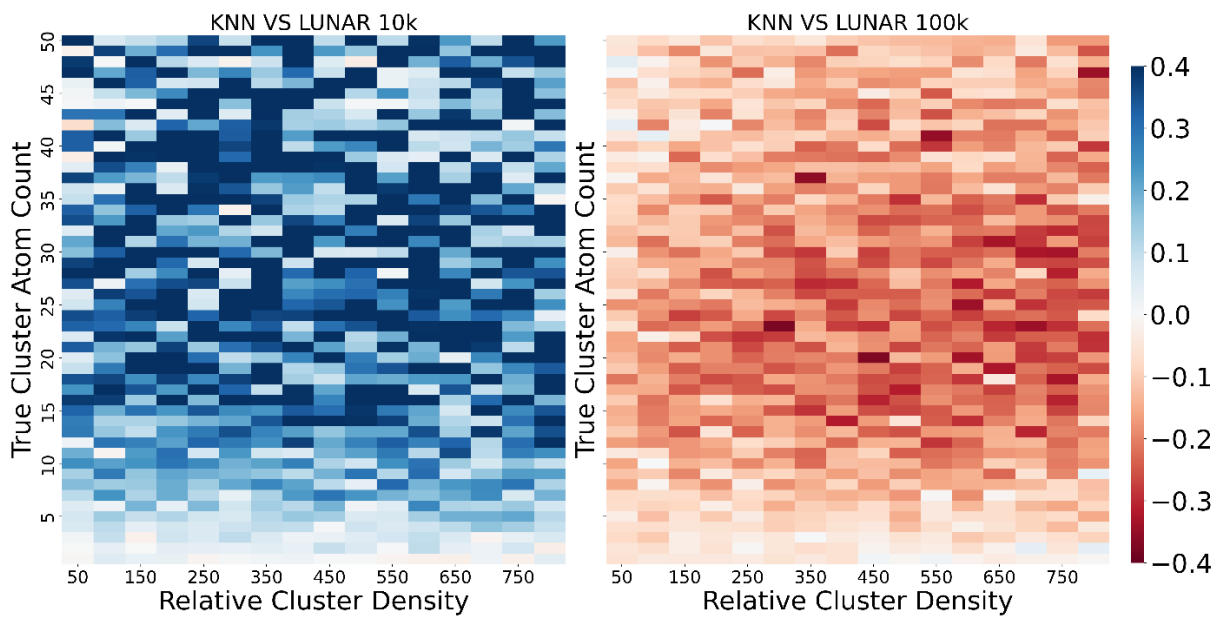

*Figure 6: Heatmaps for the comparison of kNN and LUNAR for k values of 10 and 100. For F1 score for each cluster atom count and relative cluster density pair. The negative values (red colour) correspond to LUNAR outperforming kNN, the positive values (blue colour) indicate kNN outperforming LUNAR.*

### 3.2. RPV Steel

LUNAR was used to analyse three real APT samples as outlined above, S1, S2 and S3. S1 contains three noticeable features detected by either MSM or isoconcentration surfaces: a solute cluster (seen in the bottom left of the reconstructions in Figure 7), a dislocation with large amounts of solute atom segregation (seen in the top of the reconstructions in Figure 7), and a carbide (seen in the bottom right of the reconstructions in Figure 7). All three of these features, plus an additional cluster not picked up with MSM, were found with LUNAR in a single analysis run of the APT data. This highlights one advantage of LUNAR vs. MSM, which is the ability to identify irregular shaped features, such as dislocations, carbides and grain boundaries, in addition to solute clusters. LUNAR was also able to identify a small possible cluster, seen in the centre of the reconstructions in Figure 7. This cluster was not identified by MSM.

The threshold for identifying clusters was selected using a histogram plot of the scores and visually selecting the most suitable cut-off, as seen in Figure 8. The Karcher mean was not used here due to its consistent slight underperformance relative to a manual selection. The same procedure was implemented for two further samples: S2 and S3, shown in Figure 9. S2 contains a grain boundary plus a series of clusters. S3 contains a large amount of clustering. LUNAR was able to identify the features in both samples.

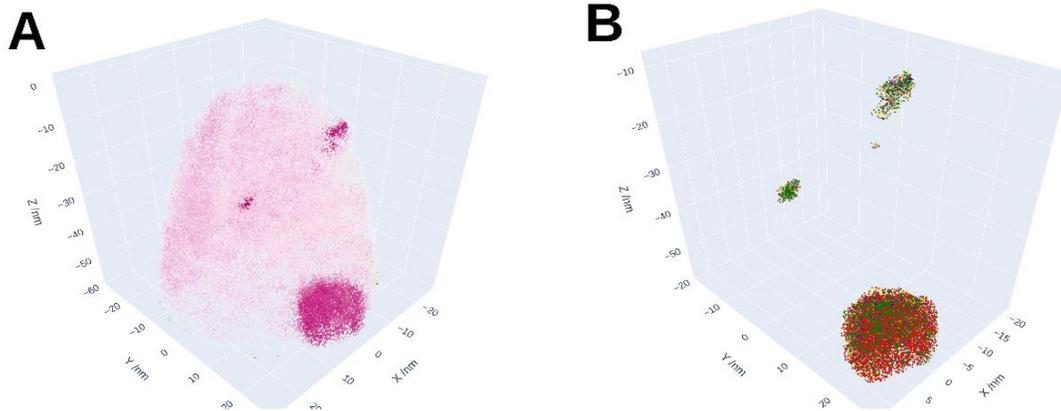

*Figure 7: A) is a 3D reconstruction of S1, an irradiated RPV steel APT dataset, that is coloured by the LUNAR output score, where dark pink corresponds to a higher clustering score against green a lower clustering score. B) is a 3D reconstruction of S1 that has been filtered according to a 12% threshold value (keeping the highest 12% scoring atoms), where the colours denote the elements; Ni-green, Mn-gold, Si-grey, Cu-orange, P-pink, Mo-red, C-brown.*

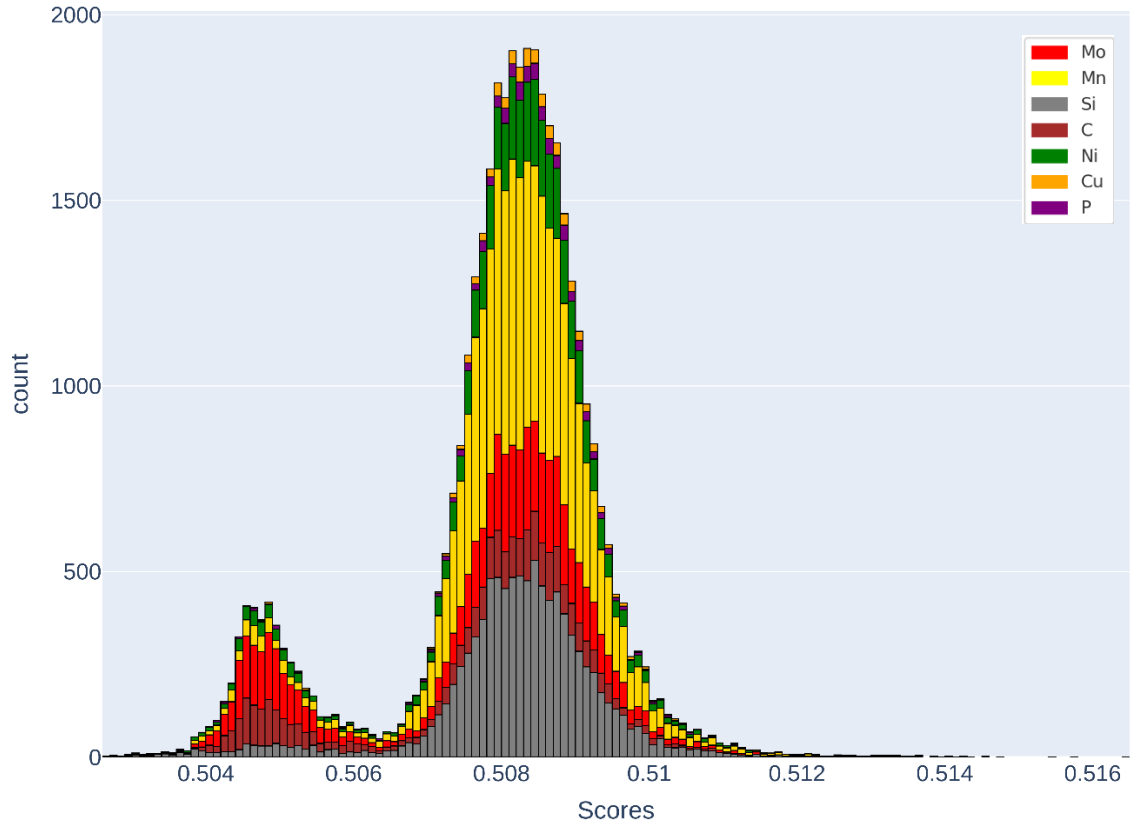

*Figure 8: Histogram of the LUNAR output scores from sample S1. The lower scored peak corresponds to the atoms segregated to the clusters and dislocations, in addition to the atoms in the carbide. The colours denote the elements; Ni-green, Mn-gold, Si-grey, Cu-orange, P-pink, Mo-red, C-brown*

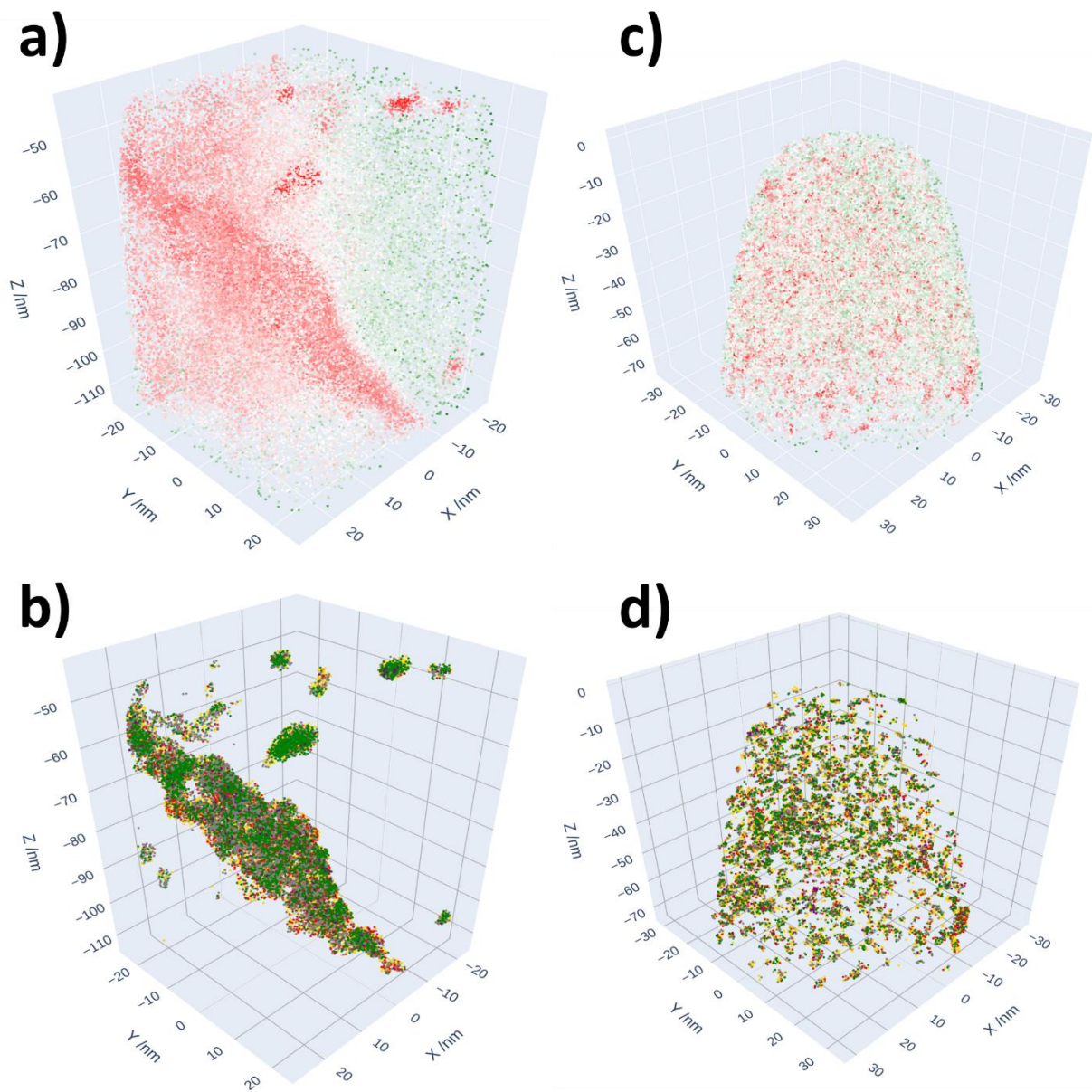

*Figure 9: a) is the unfiltered output from LUNAR for dataset S2, b) is the 12% threshold filtered output for S2, c) is the unfiltered output from LUNAR for S3, d) is the 12% threshold filtered output for S3. Where the colours denote the elements; Ni-green, Mn-gold, Si-grey, Cu-orange, P-pink, Mo-red, C-brown for b) and d). The colour scale in a) and c) correspond to the output score where red is a lower score, more likely to be a cluster, and green is a higher score.*

## 4. Discussion

The results show that neighbour-based outlier detection methods such as LUNAR have three clear advantages over the current density-based methods for cluster identification. First, LUNAR has better performance (as defined by the F1 score) for clusters with fewer than 25 atoms. Second, LUNAR with the Karcher mean thresholding cut-off has only one input parameter, the $k$ value, which mitigates human interpretation that can produce uncertainties. Third, LUNAR offers the possibility to find more irregular shaped features. With further additional statistical methods, even the $k$ input could be automated - removing all human input - such as an adaptation of silhouette analysis from [32]. LUNAR and *k*NN models were both shown to be more consistent than DBSCAN with automated thresholding and both outperformed DBSCAN for cluster sizes of fewer than 25 atoms. LUNAR and *k*NN perform similarly, with *k*NN performing slightly better on average for the simulated data. However, *k*NN appears to be more sensitive to the number of chosen nearest neighbours (Figure 6).

For dataset S3, LUNAR was able to identify the clustering present. However, the clusters are not evenly distributed throughout the sample. This is possibly a density-based artefact introduced from the APT. This may occur due to the density-based nature of LUNAR. MSM is also a density-based model which does not appear to show the same artefact. However, LUNAR's higher sensitivity may simply highlight an already existing error in the MSM outputs.

The automated thresholding technique, using the Karcher mean, consistently had the highest F1 scores among the methods tested. This may be due to the generalisation of the mean and the standard deviation to a curved geometry. This may highlight the need to consider curved geometries more closely in APT data.

LUNAR also produces a decision score, which has the potential to be used for further analysis. In particular, an exciting possibility maybe the combination of APT and atomistic modelling data. Correlative microscopy has been considered as a natural progression of APT to minimise the resolution and detection efficiency limitations. However, a *correlative* approach combining APT and atomistic modelling is seldom discussed. In atomic scale methods e.g. molecular dynamics, the forces of all of the atoms can be calculated if the atomic potential is known. However, the potential for large system is not known; it must be approximated. Several methods can be used. For example, density functional theory and machine learning approaches [33] have been used. Using experimental APT data may be able to be used to improve the approximation of the atomic potential when used in combination with other techniques. In addition, current empirical potential techniques e.g. the embedded atom method potentials are limited to just a few elements, in this case they generally only work well for metallic elements for which they were designed. For example, in RPV steel, the Bonny embedded atom potential for Fe, Mn, Ni, and Cu is commonly used [34]. Incorporating the atomic scores from an experimental APT dataset may also aid in adding more elements to atomistic simulations.

LUNAR and *k*NN both demonstrate good performance at cluster sizes above 30. This makes LUNAR or *k*NN, in standalone form, an option for a drop-in replacement for all use cases in APT data clustering detection. However, the hybrid *k*NN-DBSCAN method described may be a more suitable replacement, as shown by the results in Figure 3. It should be noted that although the hybrid method's F1 score curve is considerably higher than DBSCAN, this test was conducted on a simulated dataset without some APT specific artefacts. It is known that the density of clusters can change upon field evaporation. This effect was not included in the simulated data. For each run a cluster size was pre-determined where all of the clusters were set to the same size. If the clusters, as would more often be the case in a real dataset, were a range of different sizes, this may cause the hybrid model to revert back to DBSCAN and miss the smaller clusters. There might be a possibility for future research to consider a sampling-based approach to prevent this reversion to DBSCAN. It is likely that further improvements to the hybrid model can be made using ensemble techniques, such as voting or stacking. Ensemble methods combine multiple models together, for voting this is achieved by each model votes for a label. The majority vote determines the chosen label. For stacking, multiple models produce outputs, which are then used as inputs for another final meta-model then determines the final label.

Improvements in accuracy for smaller clusters open the possibilities to study in further detail the early formation of solute clusters. Currently, research has tended towards the study of late-stage clusters. Unfortunately, this focus does not provide answers on how clusters are formed. In fact, much of the knowledge of cluster formation comes from theory, modelling and taking larger scale clusters, whilst assuming that their formation is a scaled down version of the visible cluster. This focus on large, late-stage clusters is natural due to the present inability to detect and analyse small, early-stage clusters.

LOF is a commonly used algorithm in outlier detection. Indeed, LUNAR is in part based on LOF. However, here, LOF was unable to be adapted to identify clusters in APT. The procedure for $k$NN was identical to LOF by taking the negative of the decision score. However, LOF failed to identify clusters at a greater accuracy than random chance.

Multiple studies have highlighted the variance in analysis outcomes between different researchers in APT round robin analyses[4], [10]. This is one of the key limitations of APT. The models presented in this paper can address this by removing all human intervention after reconstruction. A future study may also reanalyse existing datasets in bulk and in a repeatable way, without human bias, allowing for a truly unbiased comparison between different samples from different researchers.

# 5. Conclusions

1. LUNAR, $k$NN and a hybrid of $k$NN and DBSCAN all had an improved classification score of up to 15% for clusters smaller than 20 atoms in size when compared to DBSCAN alone on the simulated dataset.
2. LUNAR was also used to analyse three real APT datasets, which contain a variety of features to test the model's general performance. Visually, LUNAR performed well. However, unlike with the simulated dataset, the true labels are unknown, and it is not possible to say whether LUNAR or MSM performs better.
3. Increased scores for $k$NN, for small cluster sizes, suggests that $k$NN should be used over MSM for samples with less clustering present. The use of $k$NN for cluster analysis in irradiated RPV may help detect previously undetectable embryonic clusters and reduce the need for human input, therefore reducing error from the lower input parameter count.
4. The Karcher mean was able to identify clusters via producing a threshold when applied the outlier detection scores. In the simulated data the Karcher mean proved effective at removing an input parameter, thus removing a potential cause of bias.

# CRediT authorship contribution statement

**R S Stroud:** Conceptualization, Data curation, Investigation, Formal analysis, Writing – original draft, Visualization. **Ayham Al-Saffar:** Software, Data curation, Investigation, Formal Analysis, Writing – Review & Editing, Visualization. **Megan Carter:** Investigation. **Michael P. Moody:** Supervision. **Stella Pedrazzini:**

Supervision. **Mark R. Wenman:** Conceptualization, Formal analysis, Writing – Review & Editing, Supervision, Funding Acquisition.

## Declaration of Competing Interest

The authors declare that they have no known competing financial interests or personal relationships that could have appeared to influence the work reported in this paper.

## Acknowledgments

R S Stroud and M R Wenman are grateful for financial support from EDF Energy Nuclear Generation Ltd and the EPSRC Centre for Doctoral Training in Nuclear Energy Futures under grant number (EP/S023844/1). We would also like to acknowledge the support from Dr James Douglas and Dr Christina Hofer for their assistance with the atom probe tomography.

The research used UKAEA's Materials Research Facility and the active APT facilities at the University of Oxford, which has been funded by, and is part of, the UK's National Nuclear User Facility.

## References

[1] M. K. Miller and K. F. Russell, "Embrittlement of RPV steels: An atom probe tomography perspective," *Journal of Nuclear Materials*, 2007, doi: 10.1016/j.jnucmat.2007.05.003.

[2] T. Sawabe and T. Sonoda, "Evolution of nanoscopic iron clusters in irradiated zirconium alloys with different iron contents," *J Nucl Sci Technol*, vol. 55, no. 10, 2018, doi: 10.1080/00223131.2018.1479987.

[3] A. Lervik *et al.*, "Atomic structure of solute clusters in Al–Zn–Mg alloys," *Acta Mater*, vol. 205, 2021, doi: 10.1016/j.actamat.2020.116574.

[4] E. A. Marquis *et al.*, "A Round Robin Experiment: Analysis of Solute Clustering from Atom Probe Tomography Data.," *Microscopy and Microanalysis*, vol. 22, no. S3, pp. 666–667, Jul. 2016, doi: 10.1017/S1431927616004189.

[5] G. R. Odette, T. Yamamoto, T. J. Williams, R. K. Nanstad, and C. A. English, "On the history and status of reactor pressure vessel steel ductile to brittle transition temperature shift prediction models," *Journal of Nuclear Materials*, vol. 526, p. 151863, Dec. 2019, doi: 10.1016/J.JNUCMAT.2019.151863.

[6] Y. Hashimoto, A. Nomoto, M. Kirk, and K. Nishida, "Development of new embrittlement trend curve based on Japanese surveillance and atom probe tomography data," *Journal of Nuclear Materials*, vol. 553, p. 153007, Sep. 2021, doi: 10.1016/J.JNUCMAT.2021.153007.

[7] J. Wang, D. K. Schreiber, N. Bailey, P. Hosemann, and M. B. Toloczko, "The Application of the OPTICS Algorithm to Cluster Analysis in Atom Probe Tomography Data," *Microscopy and Microanalysis*, vol. 25, no. 2, pp. 338–348, Apr. 2019, doi: 10.1017/S1431927618015386.

[8]     J. M. Hyde *et al.*, "Analysis of Radiation Damage in Light Water Reactors: Comparison of Cluster Analysis Methods for the Analysis of Atom Probe Data," *Microscopy and Microanalysis*, vol. 23, no. 2, pp. 366–375, Apr. 2017, doi: 10.1017/S1431927616012678.

[9]     S. Shah *et al.*, "Effect of cyclic ageing on the early-stage clustering in Al–Zn–Mg(-Cu) alloys," *Materials Science and Engineering: A*, vol. 846, p. 143280, Jun. 2022, doi: 10.1016/J.MSEA.2022.143280.

[10]    F. Exertier *et al.*, "Atom probe tomography analysis of the reference zircon gj-1: An interlaboratory study," *Chem Geol*, vol. 495, 2018, doi: 10.1016/j.chemgeo.2018.07.031.

[11]    Y. Wei, B. Gault, R. S. Varanasi, D. Raabe, M. Herbig, and A. J. Breen, "Machine-learning-based atom probe crystallographic analysis," *Ultramicroscopy*, vol. 194, pp. 15–24, Nov. 2018, doi: 10.1016/J.ULTRAMIC.2018.06.017.

[12]    R. A. Bennett, A. P. Proudian, and J. D. Zimmerman, "Cluster characterization in atom probe tomography: Machine learning using multiple summary functions," *Ultramicroscopy*, vol. 247, p. 113687, May 2023, doi: 10.1016/J.ULTRAMIC.2023.113687.

[13]    F. Seide, G. Li, X. Chen, and D. Yu, "Feature engineering in Context-Dependent Deep Neural Networks for conversational speech transcription," *2011 IEEE Workshop on Automatic Speech Recognition and Understanding, ASRU 2011, Proceedings*, pp. 24–29, 2011, doi: 10.1109/ASRU.2011.6163899.

[14]    J. T. Huang, J. Li, D. Yu, L. Deng, and Y. Gong, "Cross-language knowledge transfer using multilingual deep neural network with shared hidden layers," *ICASSP, IEEE International Conference on Acoustics, Speech and Signal Processing - Proceedings*, pp. 7304–7308, Oct. 2013, doi: 10.1109/ICASSP.2013.6639081.

[15]    J. Kolbusz, P. Rozycki, and B. M. Wilamowski, "The study of architecture MLP with linear neurons in order to eliminate the 'vanishing gradient' problem," *Lecture Notes in Computer Science (including subseries Lecture Notes in Artificial Intelligence and Lecture Notes in Bioinformatics)*, vol. 10245 LNAI, pp. 97–106, 2017, doi: 10.1007/978-3-319-59063-9_9/FIGURES/7.

[16]    S. Hochreiter, "The Vanishing Gradient Problem During Learning Recurrent Neural Nets and Problem Solutions," *https://doi.org/10.1142/S0218488598000094*, vol. 6, no. 2, pp. 107–116, Nov. 2011, doi: 10.1142/S0218488598000094.

[17]    F. Angiulli and C. Pizzuti, "Fast outlier detection in high dimensional spaces," in *Lecture Notes in Computer Science (including subseries Lecture Notes in Artificial Intelligence and Lecture Notes in Bioinformatics)*, 2002. doi: 10.1007/3-540-45681-3_2.

[18]    M. M. Breuniq, H. P. Kriegel, R. T. Ng, and J. Sander, "LOF: Identifying density-based local outliers," *SIGMOD Record (ACM Special Interest Group on Management of Data)*, vol. 29, no. 2, 2000, doi: 10.1145/335191.335388.

[19]    A. Goodge, B. Hooi, S. K. Ng, and W. S. Ng, "LUNAR: Unifying Local Outlier Detection Methods via Graph Neural Networks," *Proceedings of the AAAI Conference on Artificial Intelligence*, vol. 36, no. 6, pp. 6737–6745, Dec. 2021, doi: 10.1609/aaai.v36i6.20629.

[20] Y. Zhao, Z. Nasrullah, and Z. Li, "PyOD: A python toolbox for scalable outlier detection," *Journal of Machine Learning Research*, vol. 20, 2019.

[21] Christopher, "Understanding disentangling in β-VAE," *Osteologie*, vol. 25, no. 4, 2016.

[22] D. P. Kingma and M. Welling, "Auto-encoding variational bayes," in *2nd International Conference on Learning Representations, ICLR 2014 - Conference Track Proceedings*, 2014.

[23] F. Liu and H. O. Andrén, "Effects of laser pulsing on analysis of steels by atom probe tomography," *Ultramicroscopy*, vol. 111, no. 6, 2011, doi: 10.1016/j.ultramic.2010.12.012.

[24] M. Carter *et al.*, "On the influence of microstructure on the neutron irradiation response of HIPed SA508 steel for nuclear applications," *Journal of Nuclear Materials*, vol. 559, 2022, doi: 10.1016/j.jnucmat.2021.153435.

[25] S. Ioffe and C. Szegedy, "Batch normalization: Accelerating deep network training by reducing internal covariate shift," in *32nd International Conference on Machine Learning, ICML 2015*, 2015.

[26] S. Santurkar, D. Tsipras, A. Ilyas, and A. Madry, "How does batch normalization help optimization?," in *Advances in Neural Information Processing Systems*, 2018.

[27] D. Hendrycks and K. Gimpel, "GAUSSIAN ERROR LINEAR UNITS (GELUS)".

[28] N. Srivastava, G. Hinton, A. Krizhevsky, I. Sutskever, and R. Salakhutdinov, "Dropout: A simple way to prevent neural networks from overfitting," *Journal of Machine Learning Research*, vol. 15, 2014.

[29] R. Llugsi, S. El Yacoubi, A. Fontaine, and P. Lupera, "Comparison between Adam, AdaMax and Adam W optimizers to implement a Weather Forecast based on Neural Networks for the Andean city of Quito," *ETCM 2021 - 5th Ecuador Technical Chapters Meeting*, Oct. 2021, doi: 10.1109/ETCM53643.2021.9590681.

[30] M. A. Ganaie, M. Hu, A. K. Malik, M. Tanveer, and P. N. Suganthan, "Ensemble deep learning: A review," *Engineering Applications of Artificial Intelligence*, vol. 115. 2022. doi: 10.1016/j.engappai.2022.105151.

[31] S. Ramaswamy, R. Rastogi, and K. Shim, "Efficient algorithms for mining outliers from large data sets," *SIGMOD Record (ACM Special Interest Group on Management of Data)*, vol. 29, no. 2, 2000, doi: 10.1145/335191.335437.

[32] P. J. Rousseeuw, "Silhouettes: A graphical aid to the interpretation and validation of cluster analysis," *J Comput Appl Math*, vol. 20, no. C, 1987, doi: 10.1016/0377-0427(87)90125-7.

[33] A. P. Bartók, M. C. Payne, R. Kondor, and G. Csányi, "Gaussian approximation potentials: The accuracy of quantum mechanics, without the electrons," *Phys Rev Lett*, vol. 104, no. 13, 2010, doi: 10.1103/PhysRevLett.104.136403.

[34] G. Bonny *et al.*, "On the thermal stability of late blooming phases in reactor pressure vessel steels: An atomistic study," *Journal of Nuclear Materials*, vol. 442, no. 1–3, 2013, doi: 10.1016/j.jnucmat.2013.08.018.